\documentclass[twocolumn,showpacs]{revtex4-1}
\usepackage{subcaption} \usepackage{graphicx}
\usepackage{amsmath}

\begin{document} \bibliographystyle{revtex}

\title[Critical Coulomb Screening Length on Nuclear Pasta]{Effect of
  Coulomb Screening Length on Nuclear Pasta Simulations}

\author{P. N. Alcain, P. A. Gim\'enez Molinelli, J. I. Nichols and
  C. O. Dorso}

\affiliation{Departamento de F\'isica, FCEN, Universidad de Buenos
  Aires, N\'u\~nez, Argentina} \affiliation{IFIBA-CONICET}

\date{\today} 

\begin{abstract}

We study the role of the effective Coulomb interaction strength and
length on the dynamics of nucleons in conditions according to those in
a neutron star's crust. Calculations were made with a semi-classical
molecular dynamics model, studying isospin symmetric matter at
sub-saturation densities and low temperatures.  The electrostatic
interaction between protons interaction is included in the form of a
screened Coulomb potential in the spirit of the Thomas-Fermi
approximation, but the screening length is artificially varied to
explore its effect on the formation of the non-homogeneous nuclear
structures known as ``nuclear pasta''.  As the screening length
increases, we can see a transition from a one-per-cell pasta regime
(due exclusively to finite size effects) to a more appealing multiple
pasta per simulation box. This shows qualitative difference in the
structure of neutron star matter at low temperatures, and therefore,
special caution should be taken when the screening length is estimated
for numerical simulations.
\end{abstract}

\maketitle

\section{Introduction}\label{intro}

At densities and temperatures expected to exist in neutron star crusts
($ \rho \lesssim \rho_0$ and $T\lesssim 1.0 \ MeV$, with $ \rho_0$
denoting the normal nuclear density), nucleons form structures that
are substantially different from the ``normal'', quasi-spherical
nuclei we are familiar with.  Such structures, which have been dubbed
``nuclear pasta'', have been investigated using various
models~\cite{20, 21, 22, 23, 24, 25, 26, 27, 28, 29, 30, Dor12} which
have shown them to be the result of the interplay between nuclear and
Coulomb forces in an infinite medium.  The structure of the nuclear
pasta is expected to play an important role in the study of neutrino
opacity in neutron stars \cite{horo_lambda}, neutron star quakes and
pulsar glitches~\cite{32}. In neutron stars, apart from protons and
neutrons, there is an electron gas. This electron gas screens the
electrostatic long range proton-proton interaction. This screening
effect is often modeled within the Thomas-Fermi approximation,
according to which, the interaction between protons is a Yukawa-like
potential with a screening length $\lambda$:

\begin{equation*}
 V_{TF}(r) = \text{q}^2\frac{e^{-r/\lambda}}{r}
\end{equation*}

According to QFT calculations~\cite[pp. 175-180]{fetter}, the
screening length is $\lambda\approx100\,\text{fm}$.  For numerical
simulations, such a long range interaction poses a problem since, to
perform correct particle-based simulations, the simulation domain (or
cell) should be much larger than the length of the interaction
potentials~\cite{frenkel}. Using the correct value for $\lambda$ would
then require working with $\mathcal{O}(10^6)$ particles and it would
be computationally very exhaustive. Facing this issue, pioneering
authors~\cite{maru2012,horo_lambda} decided to work with a much
smaller $\lambda=10\,\text{fm}$, hoping to retain the main qualitative
phenomenological aspects of the system (competing interactions of
different length) but with smaller systems.  Even if they were indeed
capable of producing ``pasta-like'' structures, the particular choice
of the value for the screening length was arbitrary and based almost
solely on computational details.  Notably, this particular value of
the screening length was used by every author using a screened Coulomb
potential for particle-based simulations ever
since~\cite{26,horo_lambda,Dor12}.  This paper will work on studying
to which extent this arbitraty choice is physically relevant,
expanding previous works.

The role of the length of the screening has been narrowly explored by
several authors in other models.  For example, in a $2003$
investigation~\cite{30}, the screening effect of an electron gas on
cold nuclear structures was investigated using a static liquid-drop
model, and it was found that main effect of the gas screening was to
extend the range of densities where bubbles and clusters appear and to
reduce the range of stability of homogeneous phases.  While the
screening was found to be of minor importance, the study, being
static, didn't include any spatial or dynamical effect.  Another
$2005$ study~\cite{Maruyama-2005} used a density functional method to
investigate charge screening on nuclear structures at sub-nuclear
densities but still at zero temperature; in particular, cases with and
without screening were directly compared. The main results of the
study were nucleon density profiles used to quantify the spatial
rearrangement of the proton and electron charge densities.  Once
again, it was found that the density region in which the pasta exists
becomes broader when the Coulomb screening is taken into account,
mainly due to the rearrangement of the protons; the authors remark the
importance of extending such study to finite temperatures and with
dynamical models.

More recently, some works~\cite{horo13,nos13} began studying the
effect that Coulomb interaction has on the pasta formation using
dynamical models. The main findings are that artificial one-per-cell
pasta (\emph{pseudo-pasta}) could exist even when Coulomb interaction
was removed, and that they exist due to periodic boundary conditions
and finite size~\cite{binder}.

In a previous study, we showed that a combination of classical
molecular dynamics, fragment recognition algorithms and a set of
topological tools~\cite{Dor12} was very effective in the study of the
pasta structures. In particular, we showed that topological
observables can be used to classify the different structures into
recognizable patterns; this allows for cross-model comparison between
structures obtained with different approaches and to a quantifiable
analysis of the effect the nuclear and Coulomb energy have on the
pasta formation and properties. We will study the extent to which the
screening length $\lambda$ affects the morphology of the ground states
at zero temperature of nuclear pasta. To this effect, semi-classical
molecular dynamic simulations with a screened Coulomb potential and
values of $\lambda$ from $0$ to $50\,\text{fm}$ were performed. The
model used is described in section~\ref{cmd}, and in
section~\ref{coulomb-itm} numerical aspects of the Coulomb model used
are discussed. Topological tools were used to quantitatively analyse
the effect the nuclear and Coulomb energies have on the pasta
formation and its properties are introduced in
section~\ref{analysis}. Results are presented and discussed in
section~\ref{results}.

\section{Classical Molecular Dynamics Model}\label{cmd}
 
The model used here was and developed to study nuclear reactions from
a semi-classical, particle-based point of view~\cite{pandha}.  The
justification for using this model in stellar crust environments was
presented elsewhere~\cite{Dor12}, here we simply mention some basic
ingredients of the model.

The classical molecular dynamics model $CMD$, as introduced
in~\cite{14a}, is retrofitted with cluster recognition algorithms and
a plethora of analysis tools. It has been successfully used in
heavy-ion reaction studies to help understand experimental
data~\cite{Che02}, identify phase transitions signals and other
critical phenomena~\cite{16a,Bar07,CritExp-1,CritExp-2}, explore the
caloric curve~\cite{TCalCur,EntropyCalCur} and
isoscaling~\cite{8a,Dor11}.  Synoptically, $CMD$ uses two two-body
potentials to describe the motion of nucleons by solving their
classical equations of motion.  The potentials,developed
phenomenologically by Pandharipande~\cite{pandha}, are:

\begin{align*}
V_{np}(r) &=v_{r}\exp(-\mu _{r}r)/{r}-v_{a}\exp(-\mu_{a}r)/{r}\\
V_{nn}(r) &=v_{0}\exp(-\mu _{0}r)/{r}
\label{2BP}
\end{align*}
where $V_{np}$ is the potential between a neutron and a proton, and
$V_{nn}$ is the repulsive interaction between either $nn$ or $pp$. The
cutoff radius is $r_c=5.4\,\text{fm}$ and for $r>r_c$ both potentials
are set to zero. The Yukawa parameters $\mu_r$, $\mu_a$ and $\mu_0$
were determined to yield an equilibrium density of $\rho_0=0.16
\,\text{fm}^{-3}$, a binding energy $E(\rho_0)=16
\,\text{MeV/nucleon}$ and a compressibility of
$250\,\text{MeV}$~\cite{pandha}.

The main advantage of the $CMD$ model is the possibility of knowing
the position and momentum of all particles at all times. This allows
the study of the structure of the nuclear medium from a particle-wise
point of view.  The output of $CMD$, namely, the time evolution of the
particles in $(\mathbf{r},\mathbf{p})$, can be used as input in anyone
of the several cluster recognition algorithms that some of us have
designed for the study of nuclear
reactions~\cite{Dor95,Str97,dor-ran}.

As explained elsewhere~\cite{Lop00,Dor12,piekatesis}, the lack of
quantum effects, such as Pauli blocking, --perhaps the only serious
caveat in classical models-- ceases to be relevant in conditions of
high density and temperature (such as in heavy-ion reactions) or in
the low-density and low-temperature stellar environments, when
momentum trasfer between particles ceases to be important.

To simulate an infinite medium, systems with thousands of nucleons
were constructed using $CMD$ under periodic boundary conditions. Cases
symmetric in isospin (i.e. with $x=Z/A=0.5$, $2500$ protons and $2500$
neutrons) were constructed in cubical boxes with sizes adjusted to
have densities between $\rho=0.005 \ fm^{-3} \le \rho \le
0.08$. Although in the actual neutron stars the proton fraction is low
($x<0.5$), we chose to work with symmetric matter because that way we
could study the Coulomb term without having a symmetry term in the
energy.

\subsection{Coulomb interaction in the Model}\label{coulomb-itm}

To take into account the Coulomb interaction, which is formally of
infinite range, in molecular dynamics simulation under periodic
boundary conditions, it is necessary to use some approximation.  The
two most common approaches are the Thomas-Fermi screened Coulomb
potential (used with various nuclear models, e.g., in $QMD$~\cite{26},
$CMD$~\cite{Dor12} and $SSP$~\cite{horo_lambda}) and the Ewald
summation procedure~\cite{wata-2003}.  Theoretical estimations for the
screening length $\lambda$ are $\lambda\sim100\,\text{fm}$, but in the
previously mentioned works, due to computational limitations, a value
of $\lambda=10\,\text{fm}$ was chosen. Our goal on this work is to
understand the effect this \textit{a priori} arbitrary choice has on
the properties of the ground states of Neutron Star Matter within the
framework of CMD.

In this work, we used values of $\lambda$ ranging from
$\lambda=0\,\text{fm}$ (formally, no Coulomb interaction) and
$\lambda=20\,\text{fm}$, for densities $\rho=\{0.005\,\text{fm}^{-3},
0.03\,\text{fm}^{-3}, 0.05\,\text{fm}^{-3}, 0.08\,\text{fm}^{-3}\}$,
and the cut-off length was chosen at $r_c=\lambda$.  In particular for
$\rho=0.005\,\text{fm}^{-3}$, where ``gnocchi'' are formed, we
extended the analysis to $\lambda=30\,\text{fm}$ and
$\lambda=50\,\text{fm}$ to perform a quantitative analysis on the
physical properties of the clusters.

\subsection{Simulation procedure}\label{procedure}

The trajectories of the nucleons are then governed by the
Pandharipande and the screened Coulomb potentials. The nuclear system
is cooled from $T=1.6\,\text{MeV}$ to $T=0.001\,\text{MeV}$ using
isothermal molecular dynamics with the Nos\'e-Hoover thermostat
procedure~\cite{nose-hoover}, in the LAMMPS
package~\cite{LAMMPS}. Systems are cooled in small temperature s
($\Delta T\approx0.02$), decreasing the temperature once both the
energy and the temperature are stable.

\subsection{Analysis tools}\label{analysis}

The first of the analysis tools used is the pair correlation function,
$g(r)$, which gives information about the spatial ordering of the
nuclear medium. In the previous study, $g(r)$ showed that nucleons in
clusters have an inter-particle distance of about $1.8\,\text{fm}$ at
all studied densities for $\lambda=10\,\text{fm}$. It is interesting
to know if the nearest neighbor distance changes with the screening
length.

Beyond local measures, the shapes of nuclear structures can be
characterized by a set of morphological and topological observables:
their volume, surface area, mean curvature, and Euler characteristic
$\chi$. These four objects comprise the ``Minkowski functionals'' and
completely describe all morphological and topological properties of
any three-dimensional object~\cite{michielsen}.  The computation of
the mean curvature and $\chi$ can be accomplished through the
Michielsen--De Raed algorithm but requires the mapping of the nuclear
clusters into a polyhedra, procedure described in~\cite{Dor12}.

In~\cite{Dor12} it was shown that generic structures, such as
``gnocchi'', ``spaghetti'', ``lasagna'' and ``crossed-lasagnas'' or
``jungle gym'' and their inverse structures (with voids replacing
particles and vice-versa), all have well defined and distinct values
of the mean curvature and $\chi$ with magnitudes dictated by the
overall size of the structure, i.e. by the number of particles used.
For structures with near-zero Euler number, which signal spaghetti,
lasagna and their anti-structures, the values of the Minkowski
functionals are sensitive to the choice of two parameters: the size
given to each particle and the size of the cells in which we partition
the space. This classification is shown in Table~\ref{tab:mink}.

\begin{table}[ht] \centering
\caption{Classification Curvature - Euler}
\begin{tabular}{c|| c | c | c} \hline & Curvature $<0$ & Curvature
$\sim 0$ & Curvature $>0$ \\
               
\hline\hline Euler $>0$ & Anti-Gnocchi & & Gnocchi \\

Euler $\sim0$ & Anti-Spaghetti & Lasagna & Spaghetti \\

Euler $<0$ & Anti-Jungle Gym & & Jungle Gym \\ [1ex] \hline
\end{tabular}
\label{tab:mink}
\end{table}

\section{Results and discussion} \label{results}

As observed in previous works~\cite{horo13,nos13}, in absence of any
Coulomb interaction (what would be equivalent to $\lambda=0$),
pasta-like structures can be seen, although only one per cell. These
\emph{pseudo-pastas} are also shaped in spheres, rods, slabs,
anti-rods and anti-spheres, just like the pasta with Coulomb
interaction. The main difference is that, without the Coulomb
interaction, we find always one structure per cell, giving the hint
that its structure is related to the periodic boundary condition
imposed on the box. The \emph{pseudo-pasta} exists due to finite size
effects and, if the box was not to exist, the solution would be an
infinite droplet. We notice, however, that when there is Coulomb
interaction, the competition between opposing interactions gives rise
to a characteristic length. At sub-saturation densities, this
competition is responsible for the pasta phases, and in the limit of
very-low densities (and no screening) it shapes the nuclei we are used
to.

By increasing the value of $\lambda$ from $0$ to $20\,\text{fm}$ we
aim to explore the transition from artificial \emph{one-per-cell}
pasta, to more realistic situations with more than one structure per
cell. Moreover, this enables us to assess the physical implications of
the arbitrary and traditional $\lambda=10\,\text{fm}$ value.

\subsection{``Critical'' Screening Length}\label{lambda_c}

A first approach to analyze the nature of the pasta obtained is given
by taking a quick glance as the pressure of the different
configurations.

\begin{figure}[h!]  \centering
\includegraphics[width=0.9\columnwidth]{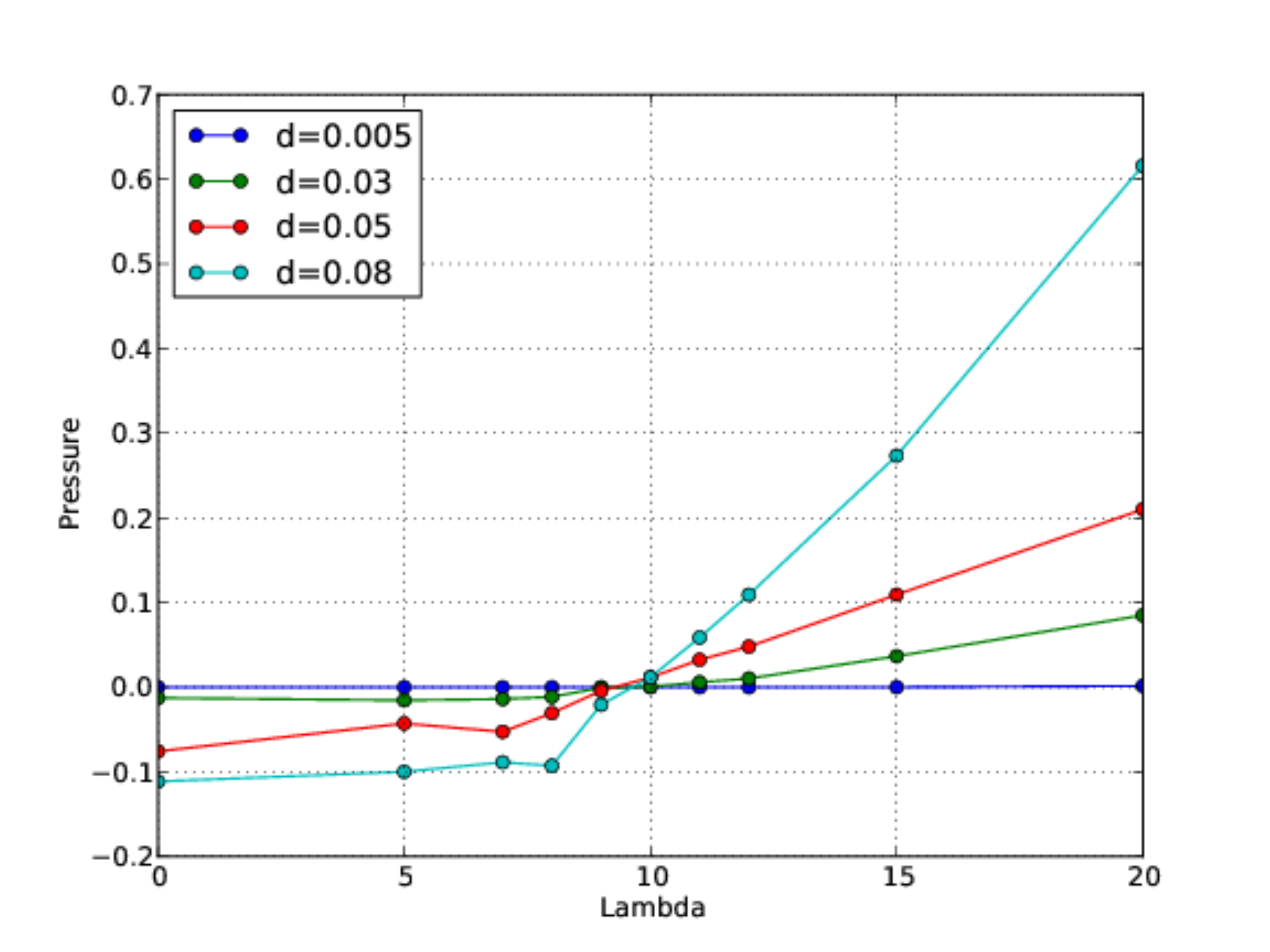}
\caption{Pressure as a function of $\lambda$ for different
  densities. We see that for $\lambda<10\,\text{fm}$, the pressure is
  negative, implying that periodic boundary conditions are affecting
  the morphology of the solution.}
\label{fig:pre}
\end{figure}

The pressure is computed by the virial formula
\begin{equation*}
P=\frac{N\,k_B\,T}{V} + \frac{1}{3}
\frac{\Sigma_{i}^{N}\mathbf{r}_i\cdot\mathbf{F}_i}{V}
\end{equation*}
where N is the number of nucleons in the system and $\mathbf{F}$ the
force exerted upon each nucleon. The terms in the virial formula
applie only to the interactions specific to the model, not
contemplating the electon gas pressure.  This pressure is not to be
mistaken with the pressure expected in neutron star crusts, since
electrons should be considered explicitly in order to calculate it
correctly, it is merely a test of the mechanical stability of the
configurations obtained within this model.  In figure~\ref{fig:pre},
we see that for all $\lambda<10\,\text{fm}$ the pressure is negative.

The negative pressure is a signal that the non-homogeneous structures
found are artificial and that the structures found can only exist
under periodic boundary conditions (see~\cite{binder,nos13}). This may
be better understood by picturing the primitive cell of the simulation
as being under the stress caused by its periodic replicas.  This mean
that for such small screening lengths the overall effective
interaction is mostly attractive and periodic boundary conditions are
still playing a major role in shaping the ground state.

For $\lambda>10\,\text{fm}$ the pressure becomes positive, meaning
that the structures formed in these configurations are not only due to
periodic boundary conditions, but Coulomb interaction is beginning to
play its intended role. The configurations for these values of
$\lambda$ indeed show density fluctuations of length smaller than the
size of the cell, which can only be attributed to the coulomb-nuclear
competition. However, the morphology of the structures, as
characterized by the topological measures described in
section~\ref{analysis}, changes drastically with $\lambda$.

\begin{figure}[h!]  
\begin{subfigure}[h!]{0.9\columnwidth}
  \includegraphics[width=\columnwidth]{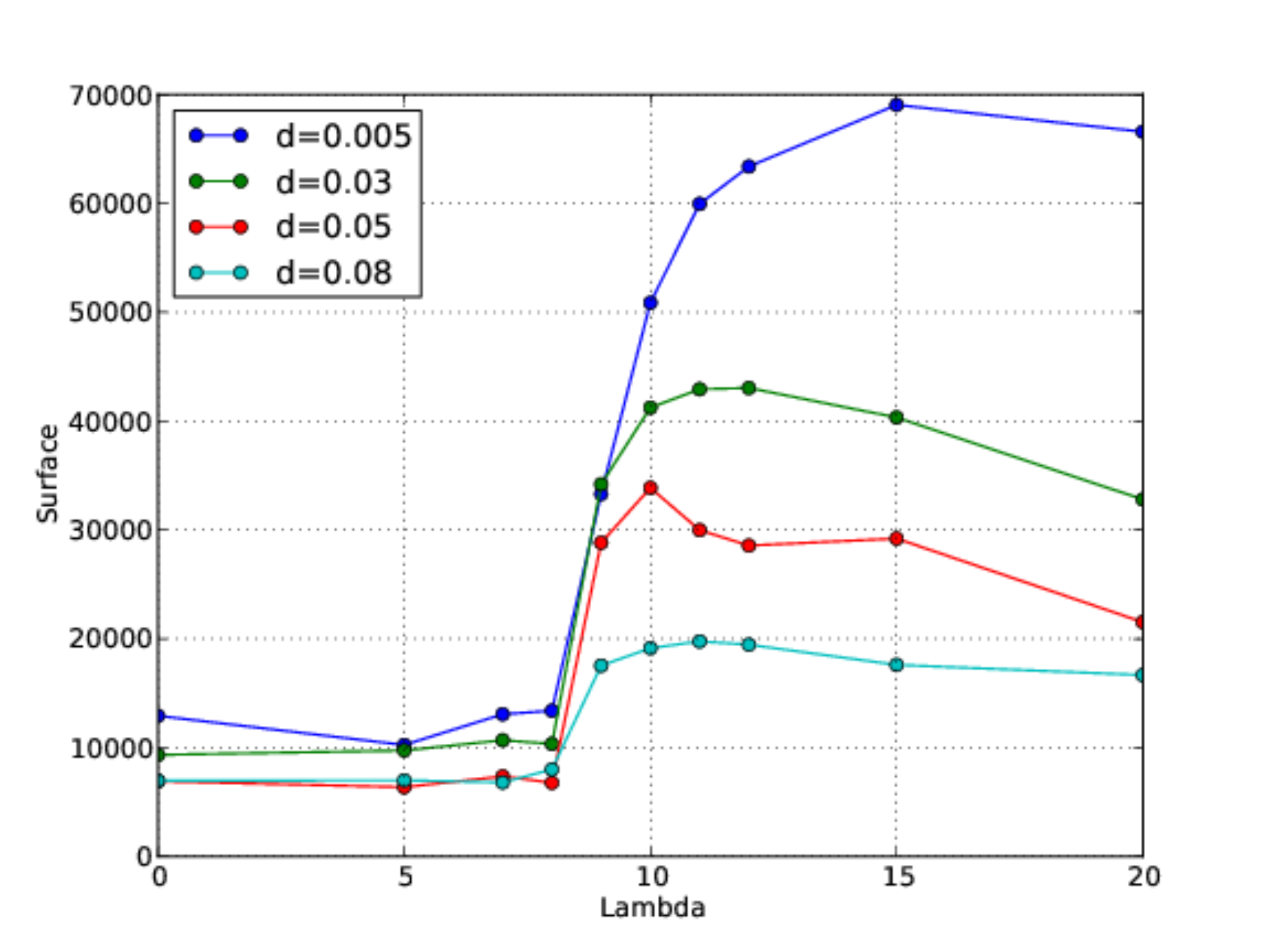}
  \caption*{Surface}
\end{subfigure}
\begin{subfigure}[h!]{0.9\columnwidth}
  \includegraphics[width=\columnwidth]{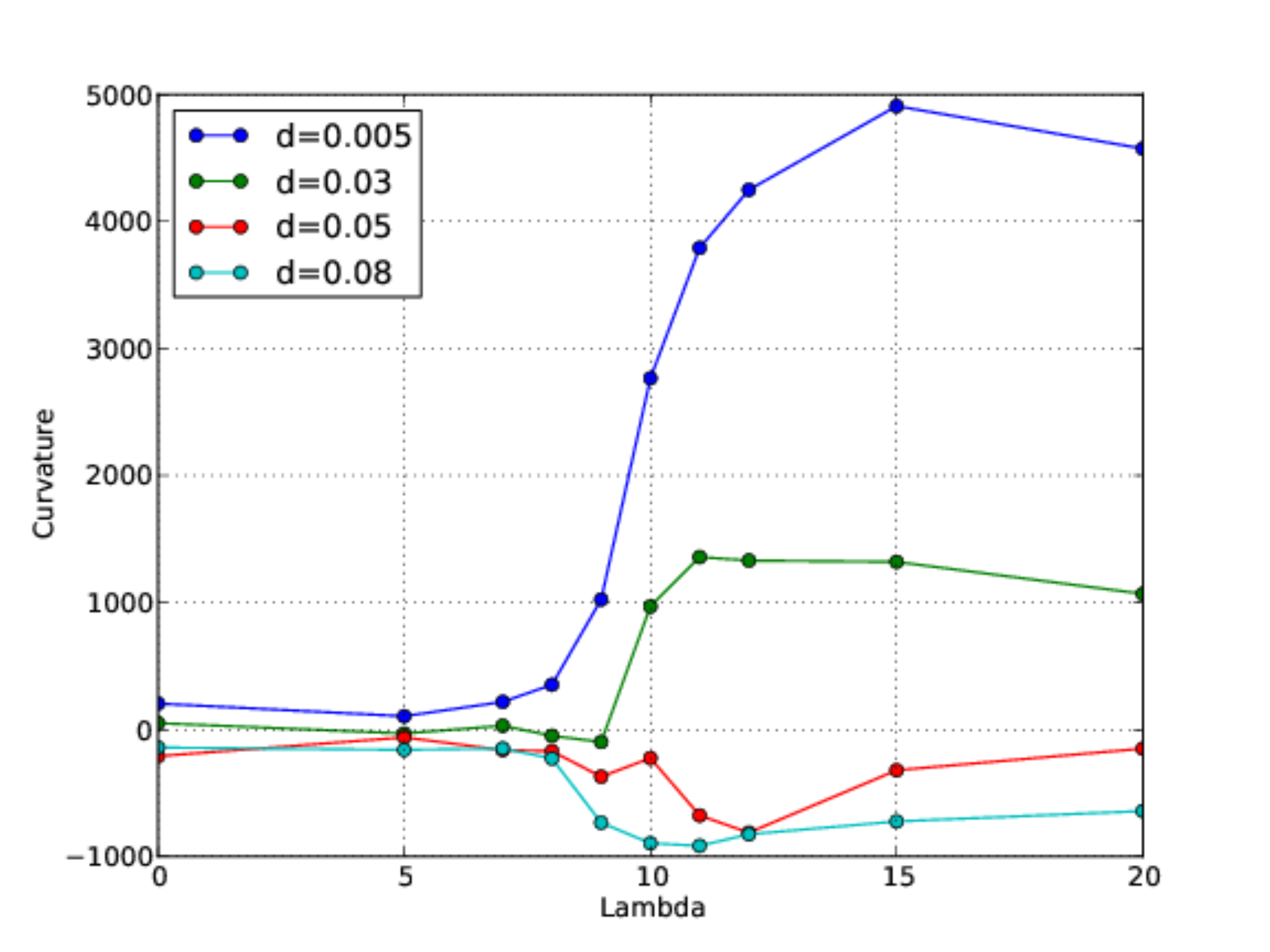}
  \caption*{Curvature}
\end{subfigure}
\begin{subfigure}[h!]{0.9\columnwidth}
  \includegraphics[width=\columnwidth]{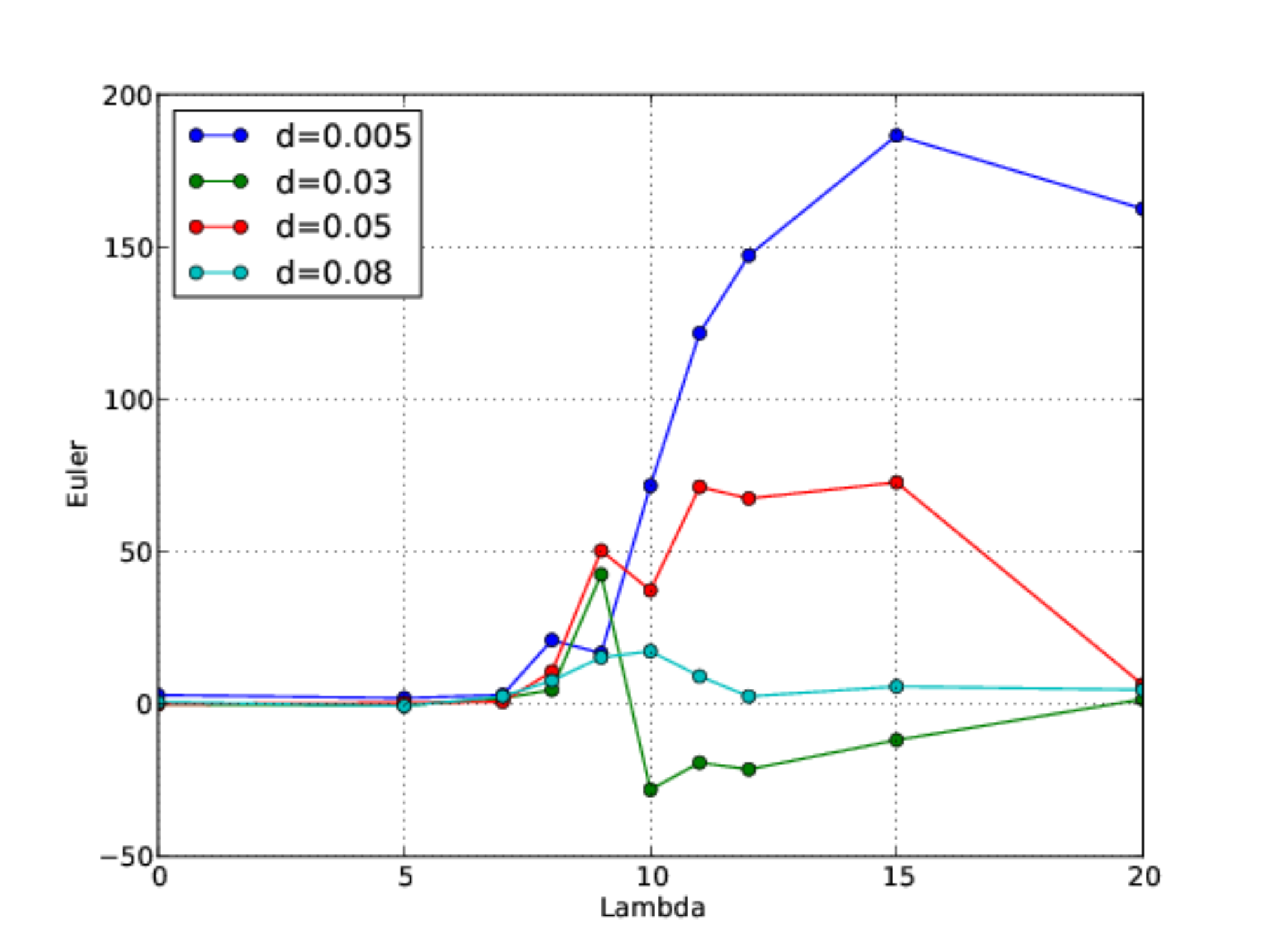}
  \caption*{Euler number}
\end{subfigure}
\caption{Minkowski functionals dependence with $\lambda$. We can see
  that there is a transition regime between $\lambda=7\,\text{fm}$ and
  $\lambda=15\,\text{fm}$, where the Minkowski functionals are
  changing.}
\label{fig:minkowski}
\end{figure}

In order to classify the low temperature ($T=0.001\,\text{MeV}$)
structures for each value of $\lambda$, we study their morphology with
the analysis tools for the spatial distribution of the particles;
namely, the Minkowski functionals. In figures~\ref{fig:minkowski} we
can see the surface, curvature and Euler number for the ground states,
and their dependence on $\lambda$ for different densities.

As stated in table~\ref{tab:mink}, we expect \emph{lasagna} and
\emph{spaghetti} to have an Euler number $\chi=0$. In the
\emph{gnocchi} case, however, each one of them contributes with
$\chi_{gn}=2$. This means that the Euler number of the whole system
will be $\chi=2\cdot\,N_{gn}$. As the configurations break up into
multiple structures per cell with increasing $\lambda$, we expect the
surface to increase as well. As for the curvature, the behavior
described in table~\ref{tab:mink} (positive for \emph{spaghetti} and
\emph{gnocchi}, zero for \emph{lasagna} and negative for
\emph{tunnel}) is only observed for $\lambda=20\,\text{fm}$. Between
$\lambda=7\,\text{fm}$ and $\lambda=10\,\text{fm}$ all three of the
Minkowski functionals change drastically before reaching well defined
values. This indicates that there is a transition regime where the
structures cannot be described as any of the traditional pasta. For
this model of nuclear interaction and Coulomb treatment, it seems the
usual $\lambda=10\,\text{fm}$ value is actually too small.

\subsection{One vs Many} \label{pasta-bup}

\begin{figure}[h!]  
\begin{subfigure}[h!]{0.45\columnwidth}
  \includegraphics[width=\columnwidth]{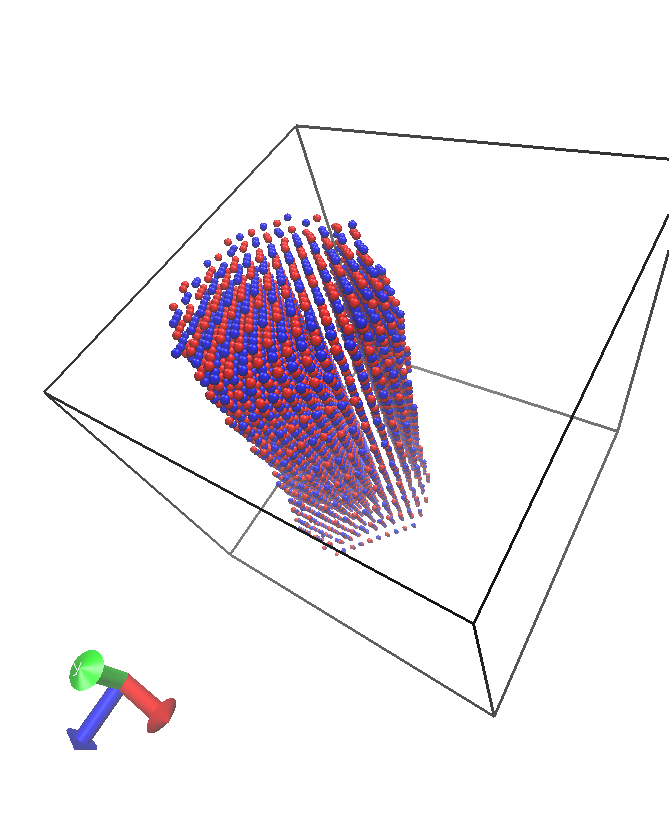}
  \caption*{$\rho=0.03\,\text{fm}^{-3}$, $\lambda=0\,\text{fm}$}
\end{subfigure}
\begin{subfigure}[h!]{0.45\columnwidth}
  \includegraphics[width=\columnwidth]{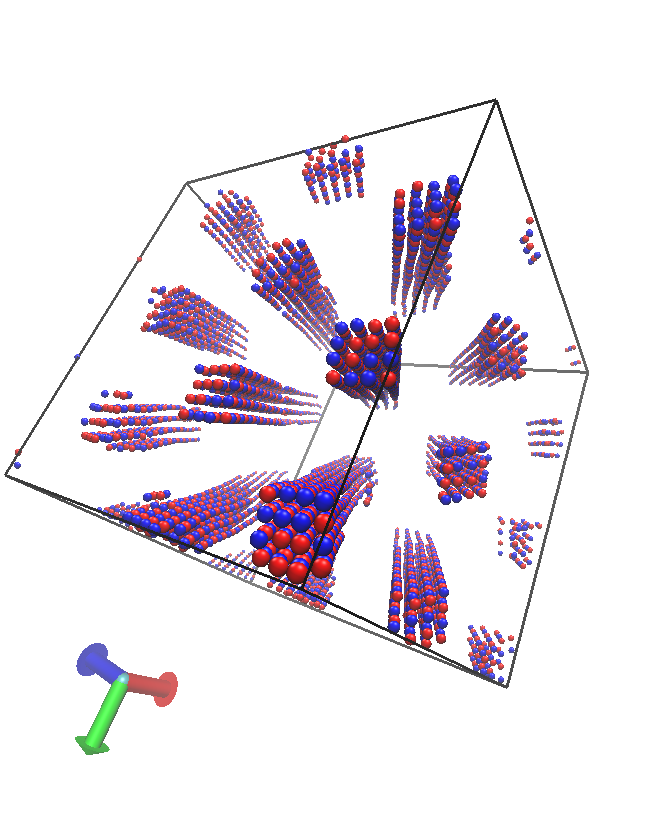}
  \caption*{$\rho=0.03\,\text{fm}^{-3}$, $\lambda=20\,\text{fm}$}
\end{subfigure}

\begin{subfigure}[h!]{0.45\columnwidth}
  \includegraphics[width=\columnwidth]{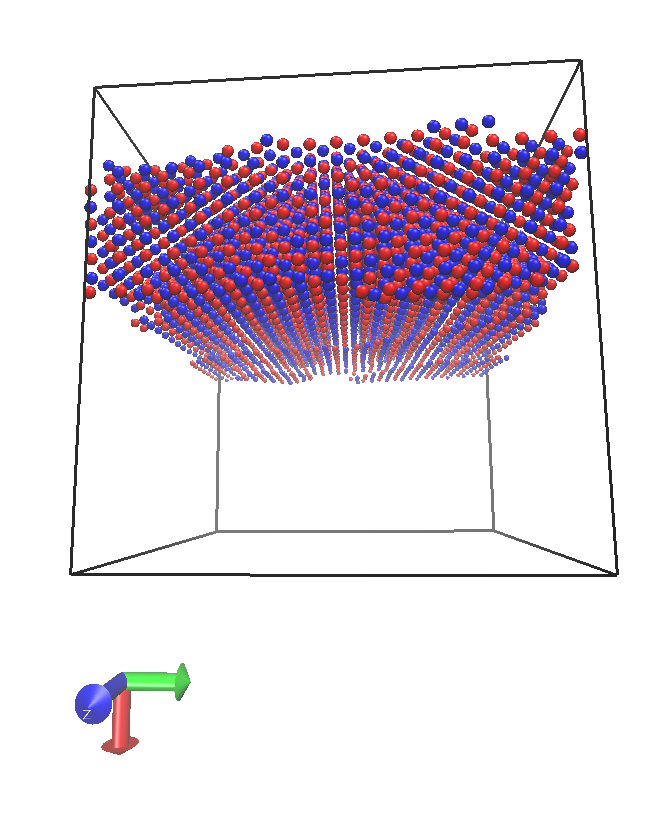}
  \caption*{$\rho=0.05\,\text{fm}^{-3}$, $\lambda=0\,\text{fm}$}
\end{subfigure}
\begin{subfigure}[h!]{0.45\columnwidth}
  \includegraphics[width=\columnwidth]{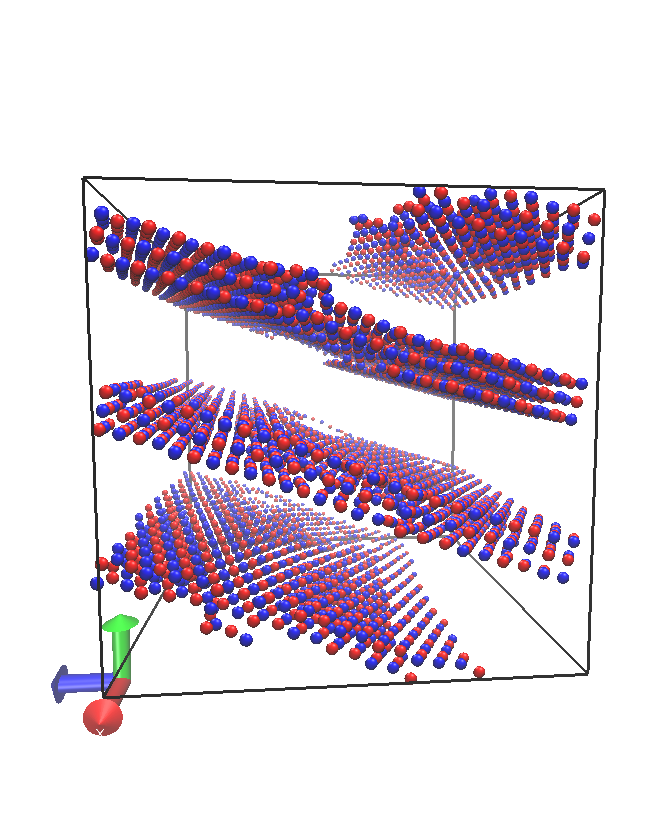}
  \caption*{$\rho=0.05\,\text{fm}^{-3}$, $\lambda=20\,\text{fm}$}
\end{subfigure}

\begin{subfigure}[h!]{0.45\columnwidth}
  \includegraphics[width=\columnwidth]{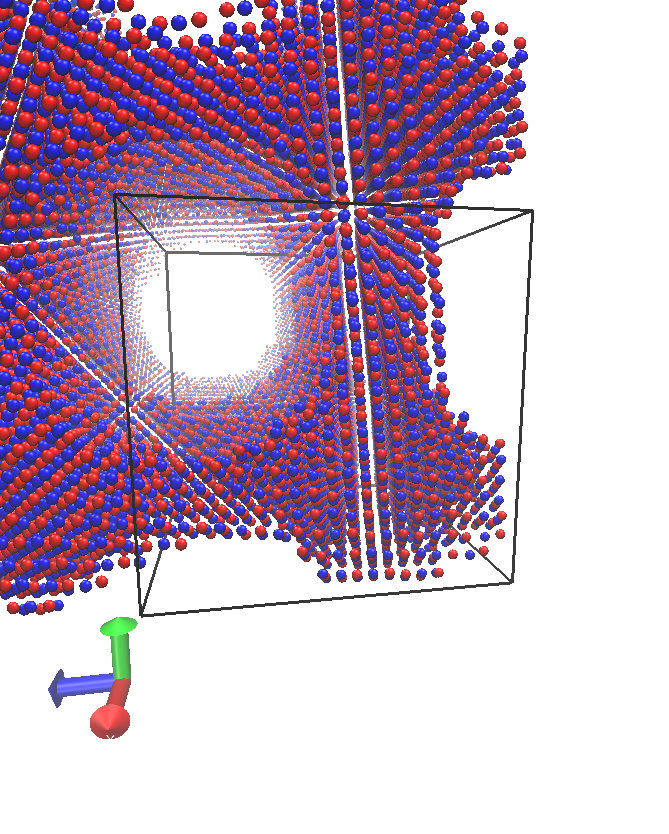}
  \caption*{$\rho=0.08\,\text{fm}^{-3}$, $\lambda=0\,\text{fm}$}
\end{subfigure}
\begin{subfigure}[h!]{0.45\columnwidth}
  \includegraphics[width=\columnwidth]{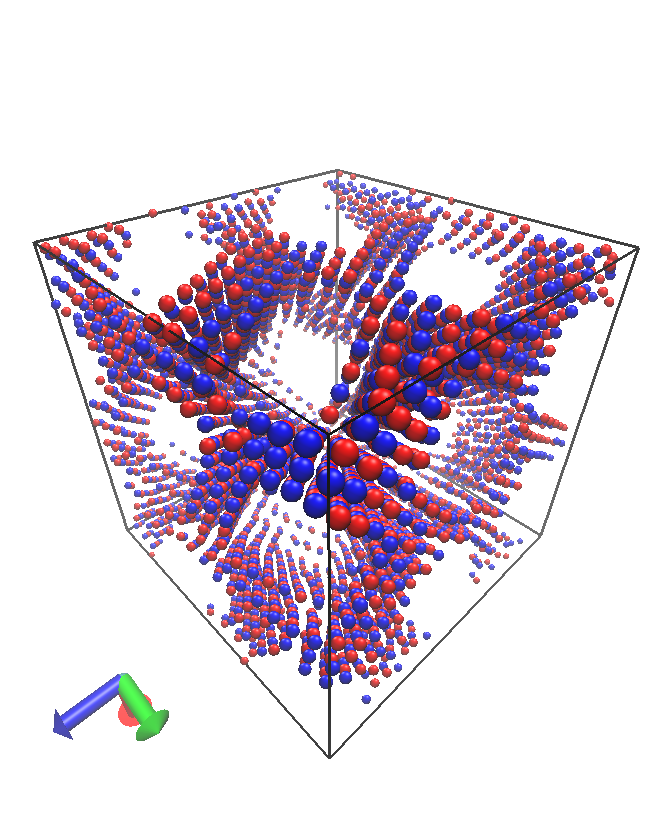}
  \caption*{$\rho=0.08\,\text{fm}^{-3}$, $\lambda=20\,\text{fm}$}
\end{subfigure}
\caption{Difference between pasta with and without Coulomb
  interaction. We can see that the Coulomb interaction splits up the
  pasta, converting one structure per cell to multiple structures per
  cell.}
\label{fig:w-wo-coulomb}
\end{figure}

To better understand how the ground state at low temperatures differ
on the transition regime from without Coulomb interaction to the
$\lambda=20\,\text{fm}$ screening length, we show in
figure~\ref{fig:w-wo-coulomb} visual representations of the results
obtained at a set of chosen densities, both with $\lambda=0$ and
$\lambda=20\,\text{fm}$.

\begin{figure}[h] 
\begin{center}
\includegraphics[width=\columnwidth]{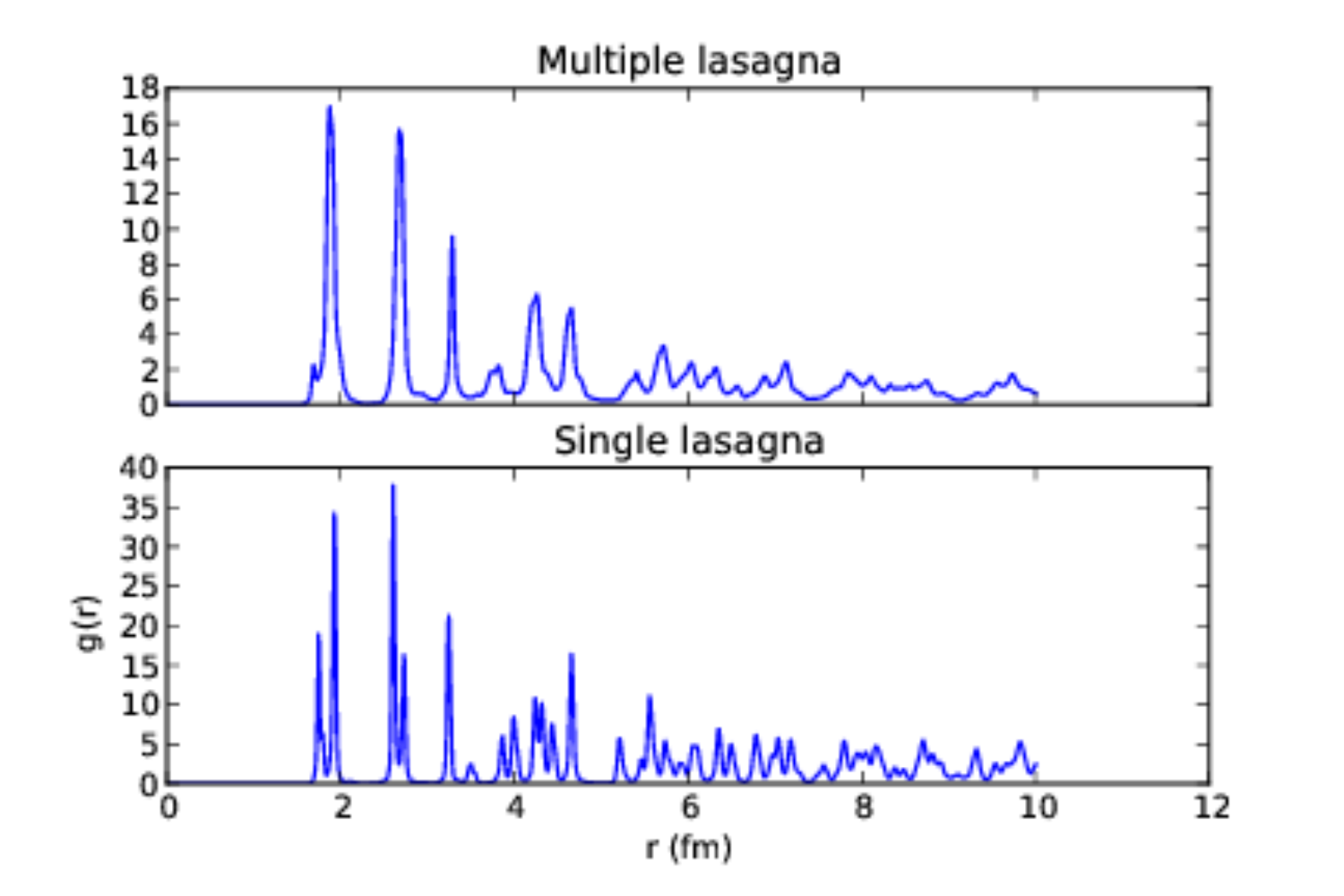}
\end{center}
\caption{Examples of the radial correlation function for
  $\rho=0.05\,\text{fm}^{-3}$ and two screening lengths:
  $\lambda=20\,\text{fm}$ (top panel) and $\lambda=0\,\text{fm}$
  (lower panel). Please notice the difference in the y-scales of the
  graphs.}\label{fig:gofr}
\end{figure}

As an example, we show the pair distribution function, $g(r)$, for
$\rho=0.05\,\text{fm}^{-3}$ in figure~\ref{fig:gofr}. In it we see
that the first peaks of the distribution remain on the same distances
of $r=1.7\,\text{fm}, 1.9\,\text{fm}$. This shows that the short range
structure is governed by the nuclear potential even at
$\lambda=20\,\text{fm}$, which is evident simply by comparing the
orders of magnitude of $V_{n-n}$ and $V_{\text{Coulomb}}$ at such
short ranges.

\begin{figure}[h] 
\begin{center}
\includegraphics[width=\columnwidth]{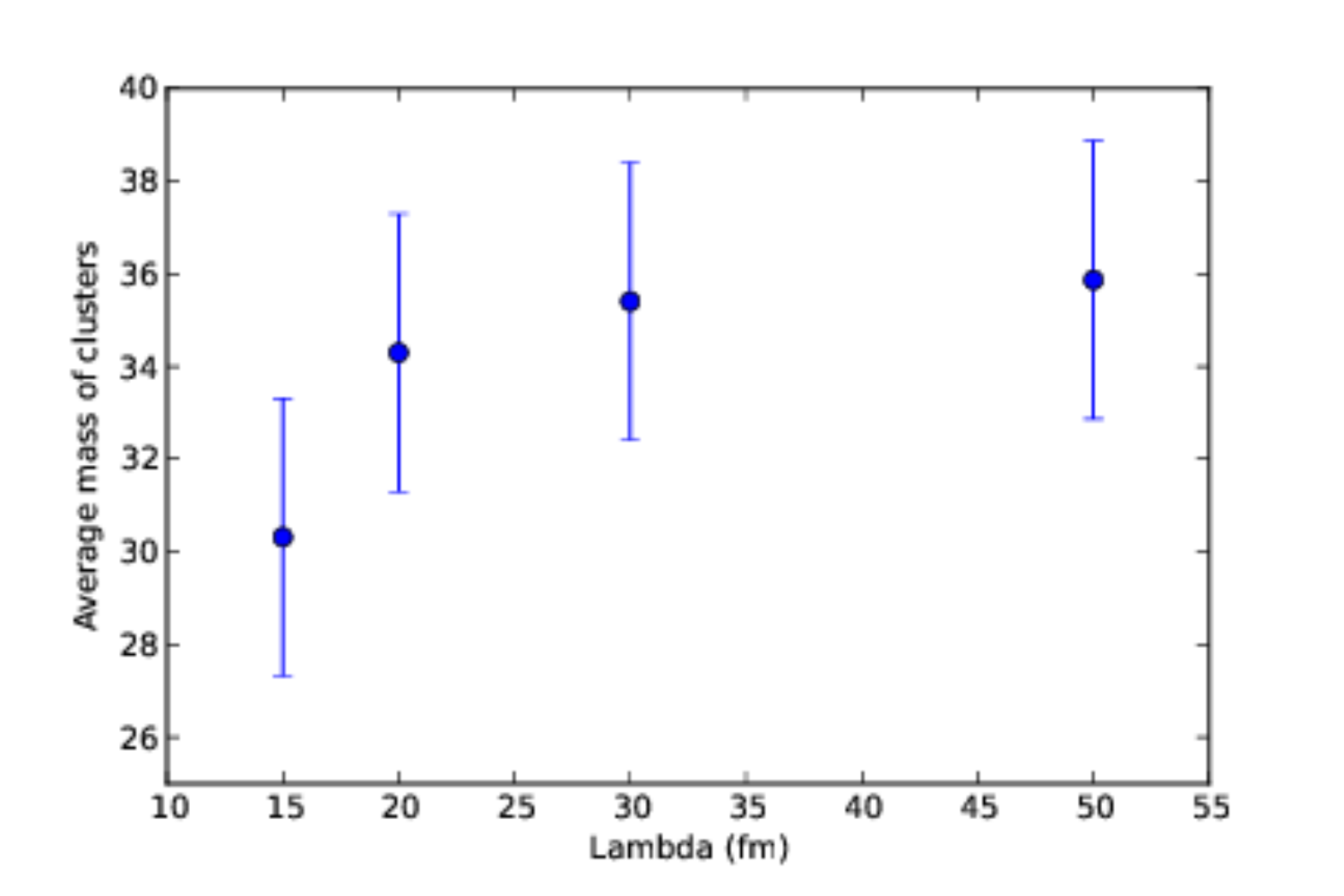}
\end{center}
\caption{Average size of nuclei depending on the screening length. We
  can see that, when considered the standard deviation, the mass
  remains the same.}\label{fig:gnocchi_mass}
\end{figure}

When increasing from $\lambda=15\,\text{fm}$ to
$\lambda=20\,\text{fm}$ at $\rho=0.005\,\text{fm}^{-3}$, although
qualitatively we see the same behavior (both show \emph{gnocchi}), the
average size of cluster is different for these two values of screening
length. This implies that the number of clusters is different, hence
the difference observed in the Minkowski functionals. To study this
result further, we plot the ``gnocchi'' size as a function of
$\lambda$ in figure~\ref{fig:gnocchi_mass}. We see that, when we
consider the standard deviation in the mass distribution, it remains
unchanged for $\lambda\geq20\,\text{fm}$. The average relative error
on this graph is $e\approx8\%$.

We can see here, however, that though the pasta structures without
Coulomb is indeed one of the known pasta structures, once we turn on
the Coulomb interaction (by making $\lambda\neq0$) the original
$\lambda=0$ \emph{pseudo-pasta} splits up: from one structure per cell
to multiple structures per cell.  For intermediate to low values of
$\lambda < 20\,\text{fm}$, the effect of the periodic boundary
conditions is still observable for some densities and more exotic
structures which can be confused with ``true'' pasta may exist.

\subsection{The Transition Regime} \label{transition}

We now turn to analyze the structures found in the transition regime.

\begin{figure}[h!]  \centering
  \begin{subfigure}[h!]{0.45\columnwidth}
    \includegraphics[width=\columnwidth]{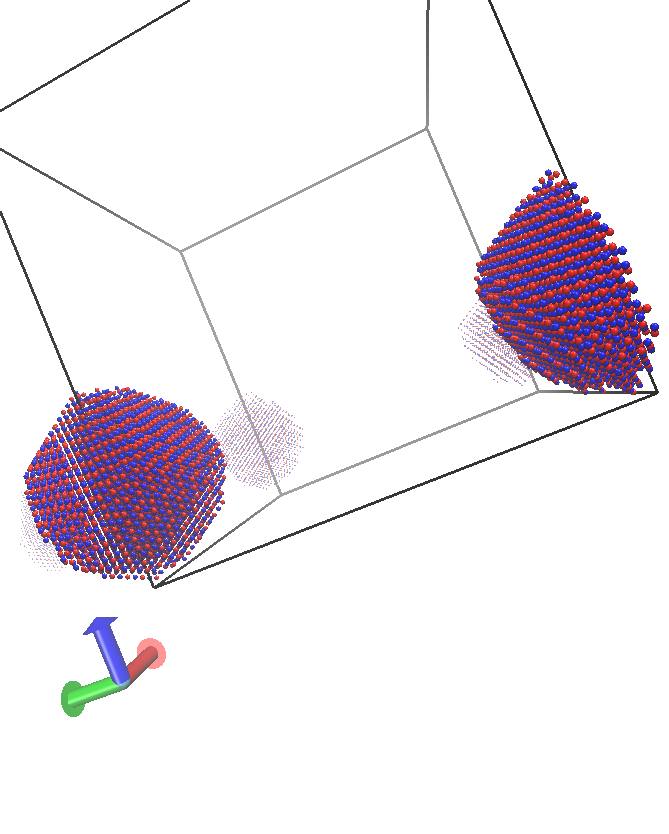}
    \caption*{$\lambda=0\,\text{fm}$}
  \end{subfigure}
  \begin{subfigure}[h!]{0.45\columnwidth}
    \includegraphics[width=\columnwidth]{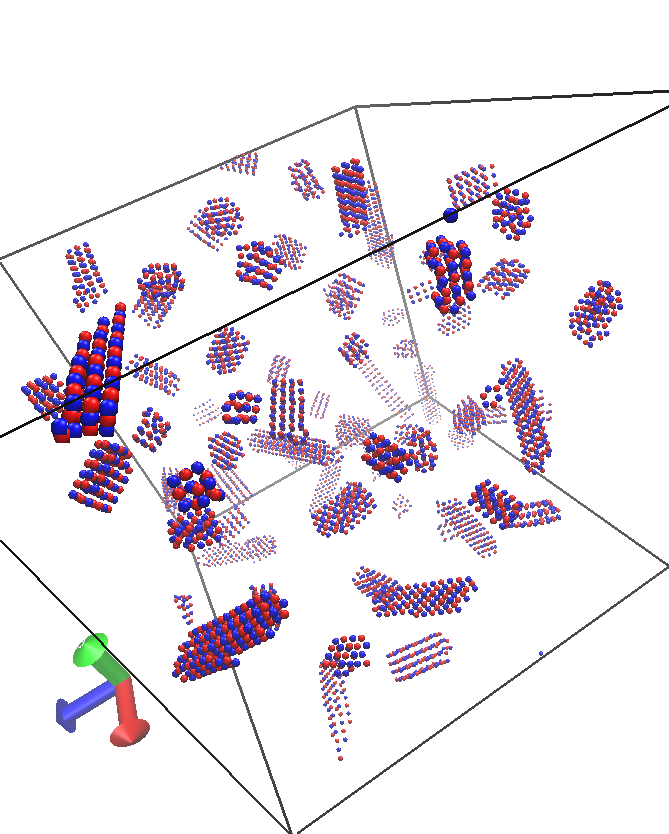}
    \caption*{$\lambda=10\,\text{fm}$}
  \end{subfigure}
  \begin{subfigure}[h!]{0.9\columnwidth}
    \includegraphics[width=\columnwidth]{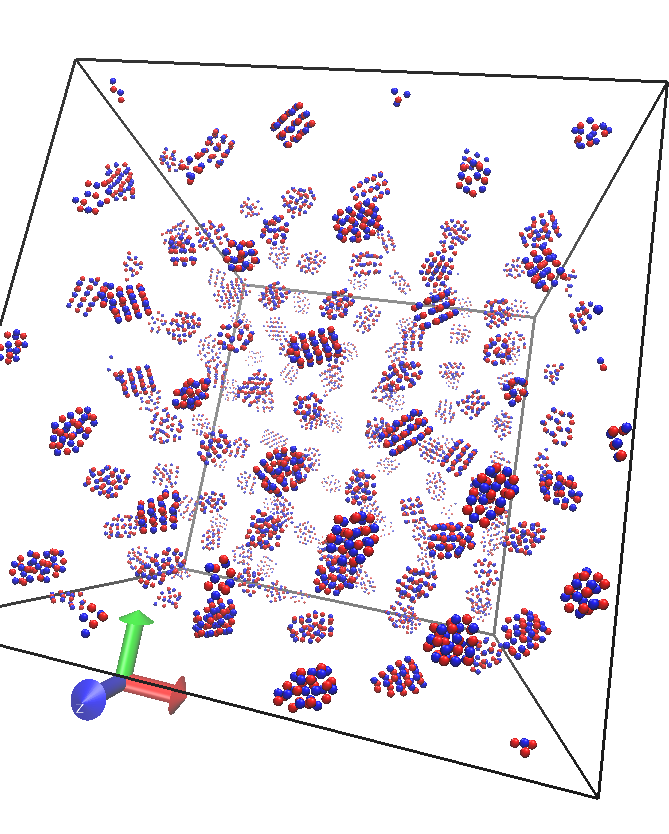}
    \caption*{$\lambda=20\,\text{fm}$}
  \end{subfigure}
  \caption{Different structures got while varying the $\lambda$
    parameter, for $\rho=0.005\text{fm}^{-3}$. In the transition
    regime, we find, at $\lambda=10\,\text{fm}$, that the structure
    breaks down to many \emph{short-spaghetti}-like parts.}
  \label{fig:gnocchi}
\end{figure}

We take, as an example, the lowest density
($\rho=0.005\,\text{fm}^{-3}$).  As can be seen in
figure~\ref{fig:gnocchi}, for $\lambda=0$ only one droplet is formed,
as expected. For $\lambda=10\,\text{fm}$, we can see that many
\emph{gnocchi} exist, but some of them stick to their neighbors
forming lumps of different sizes. Although Coulomb interaction is now
strong enough to break the one-pasta found with $\lambda=0$ into many,
the resulting droplets are not fully fledged gnocchi that can be
arranged in a regular lattice such as those for
$\lambda=20\,\text{fm}$.

\section{Discussion and concluding remarks}\label{concluding}

The effect of the screening length of the Coulomb interaction in
simulations of neutron star matter was studied at densities comparable
to that of neutron star crusts. Subsequently along the literature we
can find that the value of the screening length in the Thomas-Fermi
approximation is $\lambda\approx100\,\text{fm}$. For particle-based
simulations, due to computational limitations this value was
historically and arbitrarily reduced to $\lambda\approx10\,\text{fm}$
. This was done expecting to mantain the basic phenomenology when
simulating small sytems.  We found, though, that there is a critical
screening length $\lambda_c$ at which the structure of the ground
state drastically changes. For Pandharipande potential it lies between
$10\,\text{fm}$ and $15\,\text{fm}$ (depending on the density). For
$\lambda<\lambda_c$, the Coulomb interaction is barely acting and the
non-homogeneous structures emerging from the simulations are due to
finite size effects, as made evident from the negatve pressure of such
structures and the fact that there is only one structure per cell. For
$\lambda>\lambda_c$ the pressure becomes positive and the systems
present density fluctuations at a scale smaller than that of the cell,
but not well shaped. This transition regime is characterized by large
fluctuations in the surface, curvature and Euler characteristic $\chi$
of the structures. It is only for $\lambda=20\,\text{fm}$ that the
morphology of the structures formed stabilize and cease to deppend on
$\lambda$. Moreover, the structures in this regime are the usual pasta
phases.

Because of this, we believe extreme caution should be taken when
choosing an arbitrary value for $\lambda$, since even though some
results at $\lambda=10\,\text{fm}$ can look like the expected pasta,
the results obtained for that particular choice of $\lambda$ may be
quite different from those in the true Thomas-Fermi approximation.  In
conclusion, we find the choice of a good value for $\lambda$ that is
computationally manageable and can still adequatelly recover the
physics of the Thomas-Fermi approximation is no trivial task, and a
rigorous study needs to be done prior to the choice of the value. A
good value for $\lambda$ must lie in the $\lambda>\lambda_c$ region
which is bound to be dependent on the model used for the nuclear
interaction.

\begin{acknowledgments} C.O.D. is supported by CONICET Grant PIP0871,
P.G.M., J.I.N and P.N.A. by a CONICET scholarship. The three-dimensional
figures were prepared using the software {\it Visual Molecular
  Dynamics}~\cite{VMD}.
\end{acknowledgments}

\end{document}